\documentstyle[epsfig, twocolumn]{iso98}

\newcommand{\thepapername}{}

\newcommand{\micron}{$\mu$m}
\newcommand{\ab}{$\sim$}
\newcommand{\sun}{\hbox{$\odot$}}

\begin{document}

\setlength{\parindent}{0pt}
\setlength{\parskip}{ 10pt plus 1pt minus 1pt}
\setlength{\hoffset}{-1.5truecm}
\setlength{\textwidth}{ 17.1truecm }
\setlength{\columnsep}{1truecm }
\setlength{\columnseprule}{0pt}
\setlength{\headheight}{12pt}
\setlength{\headsep}{20pt}
\pagestyle{veniceheadings}
\markboth{Chary et al.}{Infrared Excess around White Dwarfs}

\title{\bf OBSERVATIONAL CONSTRAINTS ON THE ORIGIN OF METALS IN \\
COOL DA-TYPE WHITE DWARFS\thanks{ISO is an ESA
project with instruments funded by ESA Member States (especially the PI
countries: France, Germany, the Netherlands and the United Kingdom) and
with the participation of ISAS and NASA.}}

\author{{\bf R.~Chary, B.~Zuckerman, E.E.~Becklin} \vspace{2mm} \\
University of California, Los Angeles, USA}

\maketitle

\begin{abstract}

We present ISOCAM 7\,$\mu$m and 15\,$\mu$m 
observations of 12 nearby white dwarfs, 6 of
which have been found to have metals such as Ca, Mg and Fe in their
photospheres. Our purpose was
to search for an excess of infrared emission 
above the stellar photospheres. We find that none of the white
dwarfs other than G29$-$38 shows a detectable infrared excess and this
places strong constraints on the existence of a dusty disk around these
stars. We conclude
that ongoing accretion of the interstellar
medium seems an unlikely explanation for the existence of metals in
the photospheres of cool hydrogen atmosphere white dwarfs. 

The excess associated with G29$-$38 is
3.8$\pm$1.0~mJy and 2.9$\pm$0.6~mJy at 7\,\micron~and 
15\,\micron~respectively. The broadband spectrum of this star
strengthens the hypothesis that the infrared excess
arises from a disk of particulate matter surrounding the
white dwarf rather than from a cool brown dwarf companion.
  \vspace {5pt} \\


  Key~words: ISO; infrared astronomy; white dwarfs.
\end{abstract}

\section{INTRODUCTION}

The presence of absorption lines of Mg, Ca, Fe etc. in the
photospheres of cool (T$_{\rm eff}<20000$~K) hydrogen atmosphere DA-type white
dwarfs has been an unexplained problem for some time.  The lifetime of the
elements before they diffuse through the photosphere is in the range of a
few days, as in WD1337+70 (\cite{hol97}), 
to thousands of years 
depending on the mass and temperature of
the white dwarf. Hence, it appears as if the metals are being supplied to
the white dwarfs on relatively short timescales. The existence of metals
in the atmospheres of much hotter (T$_{\rm eff}>40000$~K) DAs has
been explained by `radiative levitation': the suspension of
metals in the photospheres by the radiation pressure of the
white dwarf (\cite{cha95}). 
However, this is not a feasible mechanism for the cooler white dwarfs
because of the diminished radiation pressure.  It has been 
convincingly demonstrated in some cases,
that the detected lines are associated with the white dwarf and not
produced by the intervening interstellar medium (ISM). For example, since
the MgII 4481\,\AA~line observed in WD1337+70 has an excited lower level,
it must be photospheric in origin (\cite{hol97}).

\cite*{dup93} showed that the metal abundances in He atmosphere white
dwarfs can be explained by episodic
accretion events whenever the white dwarf travels through relatively
overdense clouds with n$_{\rm H}$\ab10\,cm$^{-3}$ at velocities of 
$\sim$20\,km~s$^{-1}$.  However, most of the 
detected metal line white dwarfs are
within 50\,pc of the Sun and recent maps of the sodium D lines towards
stars with Hipparcos distances within 70\,pc do not show evidence for
such overdense clouds within the Local Bubble (\cite{wel98}). 

Cometary impacts every \ab10$^{4}$~yr from an Oort-like comet cloud
have also been suggested as a potential candidate for
supply of metals (\cite{alc86}). 
However, the photospheric Ca abundance in 
WD2326+049 (G29$-$38) and WD1337+70 (G238$-$44) is unlikely to be due to 
this mechanism (\cite{zuc98}). A third possibility is
accretion by mass transfer from a cool, unseen companion.

The presence of infrared excess (\cite{ZB87}) 
associated with G29$-$38 followed by the detection of metal
lines in its spectrum (Koester et al. 1997) provided the
first clues to resolve this issue. The infrared excess is thought to arise
from particulate matter orbiting the white dwarf (\cite{gra90})
although the excess might also be attributed to the presence of a cool 
companion (\cite{ZB87}).
However, speckle imaging of the source (\cite{kuc98}) 
does not indicate the
existence of a cool companion. This suggests that accretion from
the interstellar medium might be responsible for the disk of particulate
matter. 
To determine if there is a correlation between infrared excess
and the strength of metal 
lines, we undertook extremely sensitive ISOCAM mid-infrared
observations of 12 nearby ($<$25\,pc) white dwarfs, including G29$-$38, 6 of
which show metal lines in their optical/UV spectrum. 
One of these targets (WD0308+096) is a known common-envelope
binary with a red dwarf companion (\cite{saf93}).

\section{ISOCAM OBSERVATIONS AND RESULTS}
\label{sec:commands}

The ISOCAM observations (\cite{ces96}) were made in a multipoint raster
using the LW2 ($\lambda_{0}$=6.75\,$\mu$m, $\delta\lambda$=3.5\,$\mu$m)
and LW3 filters ($\lambda_{0}$=15\,$\mu$m, $\delta\lambda$=6\,$\mu$m)
with a 3 arcsec pixel field of view. 
The total integration time was of order 800\,s at LW2 and 1200\,s at LW3.
The data were mostly reduced using IDL routines developed inhouse but
the results were checked by re-analysis with the CAM Interactive
Analysis (CIA) package (\cite{ott97}). 
Photometry was performed by fitting theoretical point spread functions 
to the final reduced images.  The sensitivity of the final
map depended strongly on the number of cosmic ray hits and the nature
of the transients. However, we find that we can typically detect sources
with flux density \ab0.3~mJy at 6.75\,\micron~at the 3$\sigma$ level.
At 15\,\micron, the comparable detection threshold is about 0.5~mJy.
We also observed an A0V star HD19860, with a similar observing mode 
to obtain a calibration value from ADU/s to Jy. Using the CIA
calibration, we obtained an LW2 flux density of 19~mJy while the
predicted value for the star based on near infrared photometry was 22.1~mJy.
At LW3, the observed value was 3.4~mJy while the prediction was 4.8~mJy.
The error bars in Table 1 include this additional uncertainty in the ISOCAM
calibration.

\begin{table*}[htb]
  \caption{\em ISOCAM Observations Summary}
  \label{tab:table}
  \begin{center}
    \leavevmode
    \footnotesize
    \begin{tabular}[h]{lccccccc}
      \hline \\[-5pt]
      Object & Photospheric & Spectral Type &  \multicolumn{2}{c}{Expected
Photospheric Flux (mJy)} & \multicolumn{2}{c}{Observed Flux Density (mJy)} & Ca/H\\[+5pt]
    \cline{4-5} \cline{6-7}
       & K magnitude & &  6.75$\mu$m & 15.0$\mu$m &
LW2 & LW3 & \\[+5pt]
      \hline \\[-5pt]
        WD0046+05&11.6&DZ8 &2.4&0.5&2.0$\pm$0.4&0.5$\pm$0.2&8E-8\\
	WD0208+39&13.7&DAZ7&0.3&0.07&$<$0.6&$<$0.4&2E-9\\
	WD0308+096&15.7&DA4&0.5&0.1&0.5$\pm$0.1&....&....\\
	WD0310-68&11.9&DA3 &1.4&0.3&1.1$\pm$0.4&0.7$\pm$0.4&....\\
	WD0311-54&13.4&DZ7 &0.4&0.09&0.3$\pm$0.2&$<$0.4&....\\
	WD0322-01&14.5(J)&DZA9&0.3&0.06&$<$0.5&$<$0.4&6E-10\\
	WD0423+120&14.3(J)&DC8&0.25&0.06&0.2$\pm$0.1&$<$0.2&....\\
	WD0713+58&11.8&DA4 &1.6&0.3&1.1$\pm$0.2&$<$0.7&....\\
	WD1337+70&13.5&DAZ3  &0.3&0.07&$<$0.7&$<$0.2&2.5E-7\\
	WD1633+43&13.5&DAZ8  &0.4&0.09&0.2$\pm$0.1&$<$0.2&5E-9\\
	WD2007-30&13.0(J)&DA4&0.5&0.1&0.3$\pm$0.1&0.2$\pm$0.2&....\\
	WD2326+04&13.2&DAZ4&0.44&0.09&4.2$\pm$1.0&3.0$\pm$0.6&1E-7\\
      \hline \\
      \end{tabular}
  \end{center}
\end{table*}

\begin{figure}[h]
  \begin{center}
    \leavevmode
  \centerline{\epsfig{file=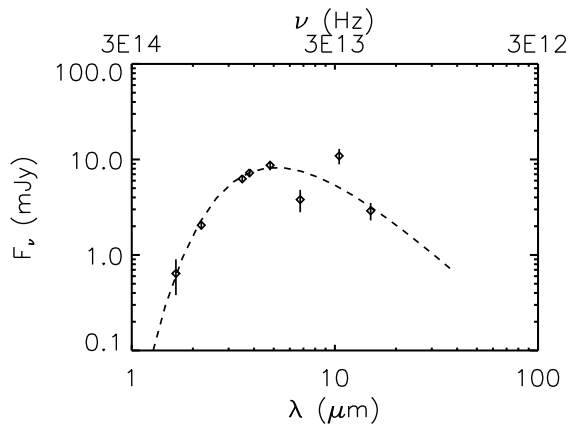,width=8.0cm}}
  \end{center}
  \caption{\em The spectral energy distribution of 
G29$-$38 after the photosphere has been subtracted. Also shown is
a 1000~K blackbody fit. Our observations strengthen the hypothesis that
the emission originates in a dusty disk rather from a cool companion.
}
  \label{fig:fig2}
\end{figure}

Table 1 shows the results of the photometry along with the predicted
values for the white dwarf photosphere based on ground based photometry 
at near infrared
wavelengths. For WD0308+096, the predicted value is the photosphere of 
the red dwarf companion which has a K magnitude of 13.15 (\cite{zb92}). 
None of the white dwarfs other than G29$-$38 shows a
significant excess of infrared emission
above the photosphere.
Also shown in the table are
derived Ca/H ratios based on calculations of
abundances from equivalent widths (\cite{koe98}, \cite{zuc98},
\cite{aan93}). Ca is a good tracer of the accretion hypothesis
since it is thought to be depleted onto dust grains in the ISM
({\cite{sav79}).

We find the excess flux density associated with G29$-$38 
at 7\,\micron~and 15\,\micron~
is substantially lower than predicted by the 800~K blackbody fit to
the excess at other wavelengths (\cite{tok90}). 
The 4.8\,\micron~flux density of the object is found to be higher in recent
IRTF data (\cite{bzl}) when compared to the initial measurement 
of \cite*{tok90}.
The value of the 4.8\,\micron~emission
after a model photosphere is subtracted is 8.7$\pm$1.0~mJy.
The 
best fit spectrum including the ISOCAM data is a \ab1000~K blackbody
which is shown in Figure 1. The deviation from the broadband spectrum 
at 7\,\micron~and 10\,\micron~might indicate line absorption/emission
processes which should be revealed in a high resolution 
mid-infrared spectrum. 

\begin{figure}[h]
  \begin{center}
    \leavevmode
  \centerline{\epsfig{file=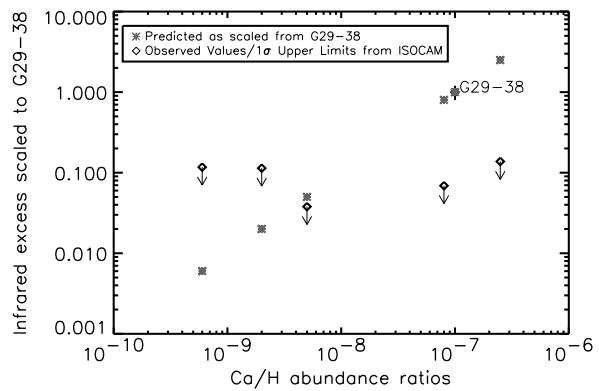,width=8.0cm}}
  \end{center}
  \caption{\em Ca/H ratios for the 6 metal line white
dwarfs against the estimates and upper limits for the excess of 
infrared emission from the ISOCAM observations.
The limits have been normalized to the excess around G29-38 which has 
a value of unity by definition.
The asterisks are estimated values for the excess scaled from the
ratio of Ca/H abundance to infrared excess for G29$-$38. 
}
  \label{fig:fig1}
\end{figure}

\section{DISCUSSION}
\label{sec:commands2}

Our mid-infrared observations were able to detect the photospheres of
most of the white dwarfs at the 2$-$3$\sigma$ level. But none of the
targets shows a significant excess above the photosphere at a level
comparable to G29$-$38 (Figure 2). 
Of the 6 white dwarfs shown in the figure, only WD0046+05 and WD1337+70 
have strong enough limits when compared with G29$-$38, to show
the lack of correlation 
between the abundance of Ca and excess infrared
emission from a disk of particulate matter.
The non-detection of infrared excess around white dwarfs
with metal abundances comparable to G29$-$38 suggests one of the following:

1. There is some mechanism whereby the dust which is responsible for
the excess of mid-infrared emission is depleted as it is
accreted onto the white dwarf. This could imply the presence of a cool,
undetected companion at a distance of about 5R$_{\sun}$ which is
clearing out the inner regions of the disk. More likely, the dust exists
as a very thin disk which absorbs less than 0.5~per~cent of the white dwarf's
radiation reducing the efficiency of thermal reprocessing. In contrast,
G29$-$38 has an
L$_{\rm excess}$/L$_{\rm WD}$ of 3~per~cent which implies a thicker 
disk possibly
due to the gravitational perturbations of an embedded object or due to
the non-radial {\it g}-mode oscillations of the star. 

2. The models for diffusion of metals through the DA white dwarf
photosphere have underestimated the diffusion timescales. This could
possibly be because of the presence of helium mixed in the
hydrogen atmosphere.
As a result, the metals might be a remnant of some episodic accretion
event in the past \ab10$^{5}$\,years.
Alternatively, the white dwarfs might have enhanced
convective layers which are dredging up the metals from some past
accretion event.

\end{document}